\documentstyle[sprocl,psfig]{article}
\def\Journal#1#2#3#4{{#1} {\bf #2}, #3 (#4)}

\def \Eslash {E \kern-.5em\slash }
\def \ccs {coupling constants }
\def \sm {Standard Model }

\def \susyq {supersymmetric }

\def\NPB{{\em Nucl. Phys.} B}
\def\PLB{{\em Phys. Lett.}  B}

\def\PRD{{\em Phys. Rev.} D}

\def\be{\begin{equation}}
\def\ee{\end{equation}}
\def\bea{\begin{eqnarray}}
\def\eea{\end{eqnarray}}

\begin{document}

\title{EXPERIMENTAL ASPECTS OF
SUPERSYMMETY WITH R-PARITY VIOLATING COUPLINGS
AT THE $e^+e^-$ LINEAR COLLIDER.}

\author{ M. BESAN\c CON }

\address{DAPNIA-CEA/Saclay, bat. 141, 91191 Gif Sur Yvette, France}

\author{ G. MOREAU }

\address{SPhT Saclay, bat. 774 \& 772, Orme des Merisiers
CEA/Saclay 91191 Gif Sur Yvette, France}


\maketitle\abstracts{A very short review on phenomenological aspects
of supersymmetry with violated R-parity at $e+e-$ colliders is given.
We present specific examples of a search for supersymmetric particles 
with violating R-parity couplings at a 500~GeV~$e+e-$ linear collider.
The signal extraction, and the mass measurement in the case of a pair production of
the lightest neutralino, 
are performed with the help of Monte Carlo simulation.}
  
\section{Introduction and Phenomenology}\label{sec:intro}
The minimal supersymmetric standard model (MSSM) 
assumes
the conservation of the leptonic ($L$) and baryonic ($B$) numbers
or, more specifically, the conservation of a so-called ($B-L$) symmetry which
can be expressed in the form of a multiplicatively conserved
quantum number, namely the R-parity $R_p = (-)^{(2j+3B+L)}$,
for a particle with spin j. The phenomenological consequences of 
the conservation of R-parity are known to be 1) the pair production
of supersymmetric particle 2) the decay of a supersymmetric particle always into
a standard particle and a supersymmetric particle 3) 
the existence of a stable Lightest Supersymmetric Particle i.e. the LSP.
The MSSM superpotential can be made more general by including
the violating R-parity $\not \! {R}_{p}$ terms \cite{rparit}:
\begin{equation}
{\lambda}_{ijk}L_iL_j{\bar E}_k + {\lambda}^{\prime}_{ijk}L_iQ_j{\bar D}_k
+{\lambda}^{\prime \prime}_{ijk} {\bar U}_i{\bar D}_j{\bar D}_k
\end{equation}
where L and E (Q and U,D) denote the left-handed component of lepton doublet
and antilepton singlet (quark doublet and antiquark singlet) chiral
superfields respectively.
The ${\lambda}_{ijk}$,
${\lambda}^{\prime}_{ijk}$
and ${\lambda}^{\prime \prime}_{ijk}$ are new Yukawa couplings,
where i,j and k are family indices going from 1 to 3.
The 9 ${\lambda}$ and 27 ${\lambda}^{\prime}$  terms violate $L$
explicitely whereas the 9 
${\lambda}^{\prime \prime}$ terms violates $B$ explicitely
\footnote{We do not discuss
the possible terms $\mu_i L_j H_2$.}.
\par
If all these new terms are present in the lagrangian, they generate
an unacceptably large amplitude for proton decay. It is therefore
generally assumed that one $\not \! {R}_{p}$ coupling dominates.
Constraints on the $\not \! {R}_{p}$ couplings 
from various experimental sources 
already exist \cite{limi}. 
One of the main
phenomenological consequence of the violation of $R_p$ is the decay of the
LSP. If the lightest neutralino ${\tilde {\chi}}_1^o$ is the LSP,
then it can
decay into a pair of fermion virtual-sfermion which sfermion decays in turn
via one of the above mentioned $\not \! {R}_{p}$ terms.
\par
At $e^+e^-$ colliders,
the effect of the $\not \! {R}_{p}$ terms on the phenomenology can be classified in  
three parts 1) {\bf t-channel or u-channel exchange} of a slepton 
via $\lambda_{ijk}$ or a squark via $\lambda_{ijk}^{\prime}$ couplings giving
rise to standard model fermion-pair production 2) {\bf single production} 
of a neutralino (with a neutrino), a chargino
(with a charged lepton) or a resonant sneutrino,
all involving $\lambda_{ijk}$ couplings and also 
single production of     
a squark in $\gamma e$ interaction involving 
$\lambda_{ijk}^{\prime}$ couplings 3) {\bf effects in the decay} of 
the supersymmetric particles
produced in pair in the usual way in $e^+e^-$ interactions. These decays can
be either direct or indirect i.e. first the usual supersymmetric (cascade) decay down
to the LSP and then the $\not \! {R}_{p}$ decay of the LSP. 
In order to have decays
inside the detector, the $\not \! {R}_{p}$ couplings have to be greater than 
$10^{-5} 10^{-6}$ ($10^{-7} 10^{-8}$) for gauginos (sfermions). This leads to a
formidable enrichment 
in the diverseness of the possible experimental signatures and topologies
in all searches for supersymmetric particles. Smaller 
$\not \! {R}_{p}$ couplings allow the supersymmetric particle to fly and decay outside
the detector. 
\par
The present study is devoted to one example of a search for a singly
produced chargino with a charged lepton and to one example of a search for
pair produced LSP.  
\section{Search for singly produced chargino with a charged lepton}
The most promising way to get a direct experimental measurement
of the $\not \! {R}_{p}$ \ccs is to study the production of particles involving 
those couplings. The reason is that the rates are then directly proportionnal
to a given power of the relevant $\not \! {R}_{p}$ coupling constant.
The $\lambda_{ijk}$ couplings can enter production reactions at leptonic colliders only.
We have considered the single chargino production, namely,
$e^+ e^- \to \tilde \chi_1^{\pm} \mu^{\mp}$, which occurs via the 
$\lambda_{121}$ $\not \! {R}_{p}$  
coupling through the exchange of either a $\tilde \nu_{\mu L}$ sneutrino
in the s channel or a $\tilde \nu_{e L}$ sneutrino in the t channel.  
Due the simple kinematic of $2 \to 2 \ body$ reaction, 
the kinematical limit of the muon transverse momentum, $P_t^{max}(\mu)$, 
which could be determined experimentally, would give either a 
relation between the $\tilde \nu_{\mu}$ mass and the $\tilde \chi^{\pm}_1$ mass,
through, $P_t^{max}(\mu)=
(m^2_{\tilde \nu_{\mu}} + m^2_{\mu} - m^2_{\tilde \chi_1^{\pm}}) 
/ m_{\tilde \nu_{\mu}}$, or 
(if the $\tilde \nu_{\mu L}$ is not produced on shell: 
$\sqrt s<m_{\tilde \nu_{\mu}}$ or
$m_{\tilde \nu_{\mu}}<m_{\tilde \chi^{\pm}_1}$)
directly the $\tilde \chi^{\pm}_1$ 
mass through, $P_t^{max}(\mu)=
(s + m^2_{\mu} - m^2_{\tilde \chi_1^{\pm}} )/ \sqrt s$.
We have assumed the LSP to be the 
$\tilde \chi^0_1$ and the dominant $\not \! {R}_{p}$ coupling 
to be $\lambda_{121}$, and we have concentrated on the chargino decay: 
$\tilde \chi_1^{\pm} \to \tilde \chi^0_1 l_p \nu_p$. Since the LSP decays as, 
$ \tilde \chi_1^{0} \to  e e \nu_{\mu} \ or \ \mu e \nu_e$, 
we obtain a $4l + \Eslash$ signature.
The main source of \sm background for the $4l + \Eslash$ final state,
which is, $e^+ e^- \to Z^0 W^+ W^-$, can be reduced using some
typical kinematical cuts (see for exemple~\cite{Godbole}).
A non-physic background comes from the $Z^0$ gauge boson pair production with 
initial state radiation of a photon (ISR). 
The $Z^0$ exclusion cut as well as some cut on the transverse missing
energy together with the beam polarization effects 
can greatly suppress this contribution.  
In the following, we show how the signal can be distinguished from
the main source of \susyq background, 
namely, the LSP pair production: 
$e^+ e^- \to \tilde \chi^0_1  \tilde \chi^0_1$.
\par
Both the signal and the SUSY background have been simulated with 
the SUSYGEN event generator \cite{susygen}. 
First, we have applied some cuts on the transverse momentum distribution 
(more peaked than the whole momentum distribution due to the ISR) 
of the produced muon. Those cuts depend on the SUSY spectrum as we have 
described above. We have also demanded that the number of muons 
must be at least equal to one, which does not affect the signal.
Finally, in order to take into account the observability 
of leptons at a Linear Collider, we have used the following cuts:
$P_t(l)>3$~GeV and $\vert \eta(l) \vert <3$. 
\par
\begin{center}
\begin{figure}
\centerline{\psfig{figure=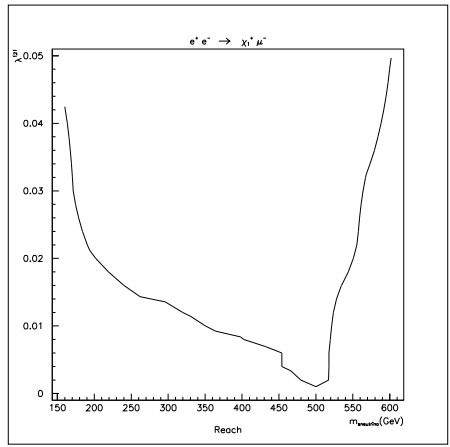,height=2.3in}}
\caption{Exclusion plot at the $5 \sigma$ level
in the plane, $\lambda_{121}$ versus $m_{\tilde \nu_{\mu}}$.
The region above the curve is the excluded region.}
\label{reach1}
\end{figure}
\end{center}

Neglecting the higgsino component of the gauginos, the incoming leptons 
in the single chargino production have the same helicity,
which is not the case for the different channels of the $\tilde \chi^0_1$ pair
production.
Hence, the beam polarization effect may be used in order to
enhance the signal-to-noise ratio. 
Furthermore, in order to reduce the SUSY background,  
we have selected the Left-handed electron and Left-handed positron in the 
initial state ($e_L^+ e_L^- \to \tilde \chi_1^{-} \mu^{+}$). 
We have assumed an electron (positron) polarization efficiency of 
$85 \%$  ($65 \%$) at the Linear Colliders.
In Fig.(\ref{reach1}), we present the exclusion plot at the $5 \sigma$ level
in the plane, $\lambda_{121}$ versus $m_{\tilde \nu}$ at a center of mass energy of
$\sqrt s=500$~GeV and assuming a luminosity of 500~fb$^{-1}$.
The chosen point of the \susyq parameter space is,
$\tan \beta =1.5$, $M_1$=150~GeV, $M_2$=300~GeV, $\mu=200$~GeV, 
$m_{\tilde l^{\pm}_p}$=300~GeV/c$^2$, $m_{\tilde q_p}$=600~GeV/c$^2$. 
The background considered here is the ${\tilde {\chi}}^0_1$ pair 
production. At the resonance point, $m_{\tilde \nu}=\sqrt s=$ 500~GeV,
the sensibility on $\lambda_{121}$ reaches, $\lambda_{121}<1.15 \ 10^{-4}$.
\section{Search for pair produced LSP}
Neutralino pairs are copiously produced at the $\sqrt s$~=~500~GeV~$e^+e^-$
linear collider. Assuming the following points 
($M_2$ in GeV, $\mu$ in GeV, $m_o$ in GeV, $\tan \beta$ )  
of the parameters space in the minimal SUGRA scheme
of the package SUSYGEN\cite{susygen} i.e. point I (200,320,100,3), 
point II (200,260160,30), point III (300,320,100,3), point IV (200,300,200,5),
the mass (in GeV/c$^2$) and total cross-section (in pb) of pair produced 
neutralino ${\tilde {\chi}}^o_1$ i.e. the LSP are, respectively, for each
point (93,0.255),   
(96,0.194), (141,0.158) and (94,0.166).
Although the $\not \! {R}_{p}$ couplings do not intervene in these LSP pair production,
we assume one of the least constrained
\cite{limi} $\not \! {R}_{p}$ couplings as the dominant 
coupling namely, the $\lambda^{\prime \prime}_{233}$ coupling which leads to a six jets
signature including 2 jets from b quark.
\par
The backgrounds from Standard Model (SM) processes are coming from 4 fermions
production such as $WW$, $ZZ$, $We\nu$ and $Zee$ as well as,
to a less extent, from 2 fermions production i.e.
two quarks production including top quark pair production.
We have also considered backgrounds coming from other 
pair produced supersymmetric particles
which are kinematically accessible at the $\sqrt s$~=~500~GeV~$e^+e^-$
linear collider such as the next to lightest neutralino ${\tilde {\chi}}^o_2$, the
lightest chargino ${\tilde {\chi}}^{\pm}_1$,
the selectron $\tilde e$, the stau $\tilde \tau$
and the sneutrino $\tilde \nu$ (for point I and II).
The LSP-signal extraction from the physical background is performed by means 
of Monte Carlo. The background events from SM processes have been generated
with the package PYTHIA\cite{pyth} and the LSP-signal events as well as
the background events from other supersymmetric particles have been generated
with the package SUSYGEN\cite{susygen}. 
\par
In order to take into account some detector effects,
we have assumed a rough smearing which includes $10 \% / \sqrt E$ 
electromagnetic resolution,
$50 \% / \sqrt E$ hadronic resolution,
90 \% electron identification probability and 80 \% efficient
tagging of event with b quark with a 1 \% contamination from $WW$ events.
\par
The LSP-signal is extracted by requiring a multiplicity of charged
particle greater than 6, a total multiplicity greater than 12. 
The total energy and the energy of the most energetic photon are required
to be respectively greater than 300 GeV and smaller than 80 GeV.
Then 6 jets are requested 
after applying a simple cluster algorithm (luclus \cite{pyth} with 
a maximum distance, below which two clusters can join, of
2.5 GeV) and each jet are required to be within
the $[10^o,170^o]$ polar angle interval. The b-tagging described above
is applied as well as a 5-constraints kinematical fit where~4~constraints
come from the 4-momentum conservation and~1~constraint comes from the requirement of 
the reconstruction
of two objects with equal invariant mass. The difference of the two 3-jets
invariant masses in the jets combination which minimizes this difference after
the above kinematical fit is required to be smaller than 15 GeV and the $\chi^2$ of
the kinematical fit is required to be smaller than 90. 
\par
\begin{center}
\begin{figure}
\begin{center}
\begin{tabular}{cc}
\psfig{figure=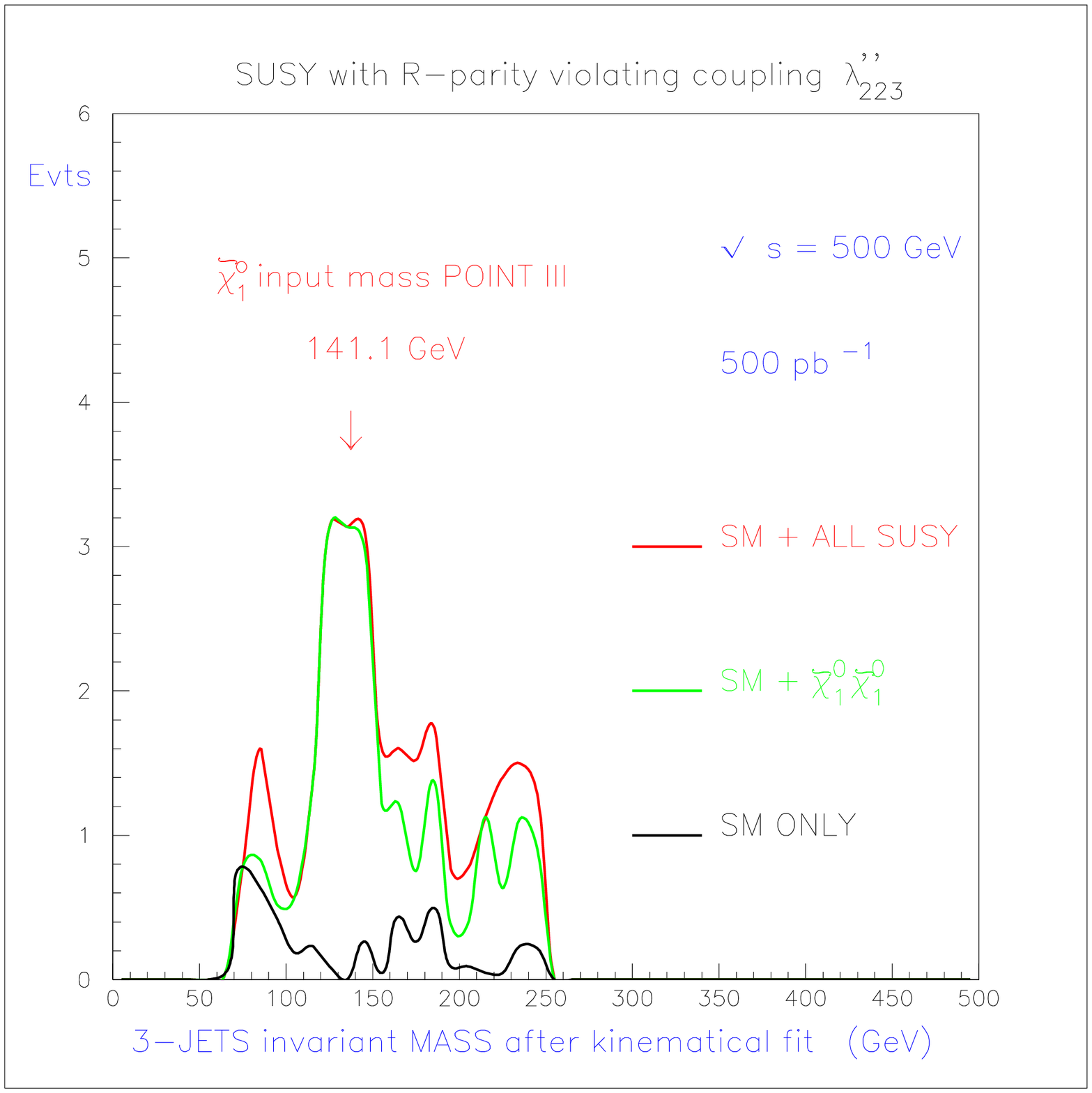,height=2.3in}
\psfig{figure=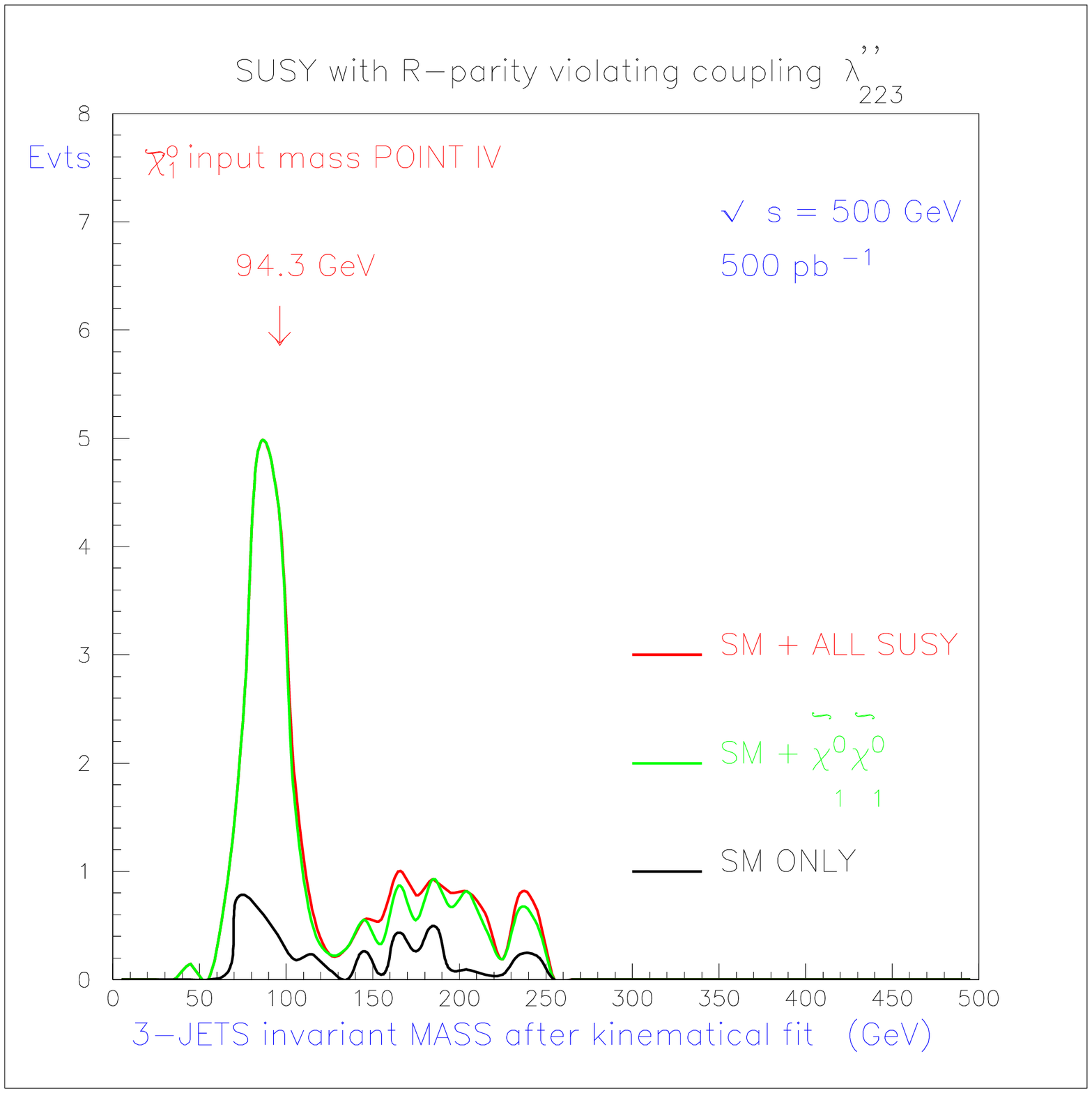,height=2.3in} \\
\end{tabular}
\caption{3-jets invariant mass after kinematical fit for point III (left figure)
and point IV (right figure).
\label{fig:fg1}} 
\end{center}
\end{figure}
\end{center}

 
This selection allows a signal efficiency in the 20 \% - 23 \% range and
a ($S/\sqrt {B}$) ratio (S for signal and B for total
background) in the 8.2 - 13.6 range for the 4 points
of the parameter space described above and with an integrated luminosity of
500 $pb^{-1}$. 
Looking at the 3-jets invariant mass after kinematical fit, the LSP mass can be
reconstructed/measured with a precision of the order of $\pm 15$~GeV to $\pm 20$~GeV,
see for example figure~\ref{fig:fg1} for point~III~and~IV. Furthermore this measurement
is not spoiled by other supersymmetric particles.
\section{Conclusion}
Supersymmetry with $\not \! {R}_{p}$ couplings leads to a very rich phenomenology
which can be explored at the 500~GeV~$e^+e^-$ linear collider. In this study, we have
presented one example of a search
for a singly produced chargino with a charged lepton in which the $\lambda_{121}$
$\not \! {R}_{p}$ coupling can be probed at the 5$\sigma$ level
to a lower value than the current low energy limit \cite{limi} 
for a wide range of the $\tilde {\nu}$ mass
and this by making use of the possibility of beams polarizations.
We have also presented a search for the 
lightest neutralino pair production where the lightest neutralino decays
via the $\lambda^{\prime \prime}_{223}$ $\not \! {R}_{p}$ coupling
leading to a 6 jets signature including 2 jets from b quark.
We have shown that for various points of the MSSM parameter space,
the signal can be extracted from the standard model processes as well as from the
production of other supersymmetric particles, which are kinematically accessible,
and the mass of the lightest neutralino can be measured. 
Such a signal extraction and mass
measurement are made more efficient (respectively) by the use of b-tagging and by means
of a simple kinematical fit.
\section*{Acknowledgments}
It is a pleasure to thank H.U. Martyn and Y. Sirois for having encouraged this work.

\end{document}